\documentclass[aps, prl, reprint, superscriptaddress, twocolumn, longbibliography]{revtex4-2}

\usepackage{natbib, color, amsfonts, amsmath, amssymb, mathtools, xr, hyperref, cleveref, graphicx, braket}
\usepackage[english]{babel}

\graphicspath{ {images/} }

\def\makeLetter{1}
\def\makeSupMat{1}

\def\metaIsPhysRev{1}
\def\metaIsAmerican{1}
\def\metaIsSyn{0}
\if\makeLetter1\if\makeSupMat1
  \def\metaIsSyn{1}
\fi\fi
\usepackage{./options}

\begin{document}
  \bibliographystyle{apsrev4-2}

  \if\makeLetter1
  \title{Compressive quantum waveform estimation}
  \author{\alextritt}
  \thanks{Correspoding author}
  
  \email{alexander.tritt@monash.edu}
  \affiliation{School of Physics \& Astronomy, Monash University, Victoria 3800, Australia.}

  \author{Joshua~Morris}
  \thanks{Current address: \emph{Vienna Center for Quantum Science and Technology (VCQ), Faculty of Physics, University of Vienna, 1010 Vienna, Austria.}}
  \affiliation{School of Physics \& Astronomy, Monash University, Victoria 3800, Australia.}

  \author{Christopher~C.~Bounds}
  \affiliation{School of Physics \& Astronomy, Monash University, Victoria 3800, Australia.}

  \author{Hamish~A.~M.~Taylor}
  \affiliation{School of Physics \& Astronomy, Monash University, Victoria 3800, Australia.}

  \author{James~Saunderson}
  \affiliation{Department of Electrical and Computer Systems Engineering, Monash University, Victoria 3800, Australia.}

  \author{L.~D.~Turner}
  \affiliation{School of Physics \& Astronomy, Monash University, Victoria 3800, Australia.}

  \begin{abstract}

    Quantum waveform estimation, in which quantum sensors sample entire time series, promises to revolution\is e the sensing of weak and stochastic signals, such as the biomagnetic impulses emitted by firing neurons.
    For long duration signals with rapid transients, regular quantum sampling becomes prohibitively resource intensive as it demands many measurements with distinct control and readout.
    In this \Letter, we demonstrate how careful choice of quantum measurements, along with the modern mathematics of compressive sensing, achieves quantum waveform estimation of sparse signals in a number of measurements far below the Nyquist requirement.
    We sense synthes\is ed neural-like magnetic signals with \acrlong*{rf}-dressed ultracold atoms, retrieving successful waveform estimates with as few measurements as compressive theoretical bounds guarantee.
  \end{abstract}

  \maketitle

  \footnotetext[0]{See Supplemental Material at \emph{[URL will be inserted by publisher]} for a derivation of sensitivity to sine coefficients, an explanation of regular\is ation parameter tuning, an explanation of how we use the \acrshort*{auc} statistic from signal detection theory\ox\ and recoveries from genuinely undersampled measurements.}
  \prSection{Introduction}%
    Quantum waveform estimation extends the applicability of precise quantum sensors to the realm of sampling entire signals~\cite{tsang_fundamental_2011}, at the cost of requiring many quantum measurements.
    One proposed application is the non-invasive sensing of electrical communication in a network of neurons by recording proximal magnetic fields~\cite{barry_optical_2016, webb_detection_2021, troise_laser_2021, webb_high-speed_2022, hansen_microscopic-scale_2022, xu_wavelet-based_2016, karadas_feasibility_2018, parashar_axon_2020}.
    Such measurements would aid in understanding the brain on the microscale, in health and in disease~\cite{mcdonald_clinical_2012}.
    Neural waveforms, composed of bipolar pulses separated by large gaps~\cite{mitterdorfer_potassium_2002}, clearly contain scant information:
    sampling one uniformly at its Nyquist rate~\cite[\Section~7.1]{oppenheim_signals_1997} will measure mostly silence.
    If information is so sparse, then could one save on resources by taking far fewer quantum measurements?
    Na\"ively reducing a uniform sample rate, however, would ultimately lose pulses to aliasing.
    In this \Letter, we use \emph{compressive sensing} to demonstrate quantum waveform estimation of a sparse, neural-like magnetic signal using many fewer quantum measurements than the Nyquist-Shannon sampling theorem would deem necessary.

    The modern mathematics of compressive sensing~\cite{donoho_compressed_2006, koep_introduction_2019, foucart_mathematical_2013} (also \emph{compressed sensing} or \emph{compressive sampling})
    describes how, under certain conditions, sparse signals can be fully recovered from an \emph{incomplete} set of measurements.
    Compressive sensors thrive where there is benefit to taking few measurements, \eg 
    cameras for wavelengths with expensive photodetectors~\cite{duarte_single-pixel_2008}, and
    \acrlong*{ct} scans which irradiate patients~\cite{graff_compressive_2015}.
    Compressive sensing in quantum science has assisted quantum process and state tomography~\cite{teo_modern_2021}, quantum communication~\cite{ran_tensor_2020, sherbert_quantum_2022}, quantum computation~\cite{fontana_efficient_2022, seif_compressed_2021, huang_random_2023}, quantum annealing~\cite{ayanzadeh_ensemble_2020, aonishi_l0_2022,ayanzadeh_quantum-assisted_2022, gunathilaka_effective_2023}, ghost imaging~\cite{duarte_single-pixel_2008, katz_compressive_2009, johnson_how_2020,cao_Zero-photon_2021}\ox\ and quantum sensor readout~\cite{arai_fourier_2015, holland_fast_2011}.
    In quantum waveform estimation, compressive sensing has been used to denoise signals~\cite{muller_nuclear_2014, puentes_efficient_2015} or disambiguate frequencies in complete sets of quantum measurements~\cite{boss_quantum_2017}.
    \citet{magesan_compressing_2013} proposed literal \emph{compressive quantum sensing}, simulating the reconstruction of a neural-like signal from an incomplete set of an incomplete set of Walsh basis measurements using sparse recovery.
    However, the proposed sensor, based on \acrfull*{nv} cent\re s in diamond, would decohere before even a single neural firing would complete\ox\ and no such sensor has been constructed.

    Compressive quantum sensors that significantly reduce the number of required quantum measurements would make quantum waveform estimation much more practical.
    In particular, it would bring into view quantum sensor arrays capturing entire waveforms in a single shot.
    To capital\is e on this promise, we need:
    (1) quantum sensors that remain coherent for the duration of the signal,
    (2) a quantum sensing protocol to measure in a basis that collects information uniformly\ox\ and
    (3) a post-processing algorithm that can robustly recover the signal from these measurements.
    Compressive sensing satisfies (3); here we demonstrate that \acrlong*{rf}-dressed cold atoms can satisfy (1) and (2). 

    Sensing electrical signals from a living \emph{network} of neurons \invitro would provide information about how nerve cells communicate on the network scale, considered crucial for understanding diseases such as epilepsy~\cite{mcdonald_clinical_2012}.
    Conventionally, neural pulses (action potentials) are measured using invasive patch clamps galvanically connected to individual neurons~\cite{mitterdorfer_potassium_2002}.
    The process perturbs neural connections and can damage cell membranes, while not scaling beyond microscopic regions of tissue.
    This motivates ongoing work to develop non-invasive methods which infer neuronal current waveforms from their corresponding magnetic fields.
    Being on the order of nanotesla at sensing range~\cite{mitterdorfer_potassium_2002, webb_detection_2021},
    quantum sensors are natural candidates for magnetometers that could sense such neuromagnetic signals.
    Quantum magnetometry of living nerve cells to date~\cite{barry_optical_2016, webb_detection_2021, troise_laser_2021, webb_high-speed_2022, hansen_microscopic-scale_2022} has been achieved with continuous-time ensemble sensing, averaged over many iterations of a nerve cell being stimulated, assumed each time to evoke an identical response.
    In contrast, we demonstrate a projective measurement protocol, which if extended to a quantum sensing array, would plausibly sense a unique neural waveform, containing multiple firings, without averaging.

    Neural waveforms are an example of \emph{sparse} signals: those with only a small proportion of non-zero values [see \Figure~\ref{fig:overview}(a)].
    Consequently, information in neural waveforms is concentrated in short windows of time, and sampling in time is inherently inefficient~\cite{mouradian_quantum_2021}.
    Any uniform quantum measurement of such sparse signals will waste quantum resources (\eg sensing duration of \acrshort*{nv}s, or a limited number of projective measurements in a sensing array) by making measurements that are not very informative.

    We instead reduce the number of measurements by taking non-uniform samples in a \acrfull*{dst}~\cite[\Equation~(37)]{jain_fast_1976} basis. 
    Time sparse signals have information spread across their broad frequency spectra.
    Hence, one can learn all information about sparse signals using only an incomplete subset of frequency measurements~\cite{donoho_uncertainty_1989}, seemingly violating the Nyquist-Shannon sampling theorem~\cite[\Section~7.1]{oppenheim_signals_1997}.
    This is the central idea of \emph{compressive sensing}~\cite{koep_introduction_2019, foucart_mathematical_2013}.
    While many possible signals are consistent with a randomly chosen incomplete set of frequency measurements of a sparse signal, it is very unlikely any will be as sparse as the true signal~\cite[\Theorem~6]{koep_introduction_2019}.
    One then recovers the true signal from incomplete measurements using an optim\is ation algorithm to find the sparsest signal consistent with the measurements and with a linear model of the sampling process.
    A hurdle for applying this to \emph{quantum} sensing is that, to measure a frequency coefficient, the sensor would need to be coherent for the duration of the experiment.
    Zeeman coherence times of several seconds make trapped cold atoms a promising candidate for such a task.

    Here, we outline how cold atoms can measure in a \acrshort*{dst} basis.
    We then describe a recovery process that retrieves neural-like signals from an incomplete set of such measurements.
    Finally, we demonstrate experimental waveform estimations derived from incomplete data, acquired with dressed cold atoms exposed to synthes\is ed neural-like signals.

  \prSection{Cold atom model}%
    We model our cold $\rubidiumLxxxvii$ atom clouds as ensembles of non-interacting spin-one systems that couple to magnetic field $\mathbf{\magneticField}(\tim)$ via $\hamiltonian(\tim) = -\gyro\,\mathbf{\magneticField}(\tim)\cdot \mathbf{\spin}$, where $\spin_{\geomX, \geomY, \geomZ}$ are hyperfine spin-1 operators and $\gyro$ is the appropriate gyromagnetic ratio.

    We aim to measure the Fourier sine coefficient $\measurement(\frequency)$ of a signal $\groundTruthSignal(\tim)$, \ie
    \begin{align}
      \measurement(\frequency) =& \frac{1}{\sensingDuration}\int_{0}^{\sensingDuration}\sin(\mtau \frequency \tim)\,\groundTruthSignal(\tim)\,\diff\tim,\label{eq:dst}
    \end{align}
    for different frequencies $\frequency$.
    To do this, we apply three fields: a bias generating Zeeman splitting by $\larmorFrequency$, transverse resonant \acrlong*{rf} dressing with Rabi frequency $\rabiFrequency$, and the signal field $\groundTruthSignal(\tim)$ parallel to the bias.
    The sensor evolves under the resulting Hamiltonian 
    \begin{align}
      \hamiltonian(\tim) = \larmorFrequency \,\spin_\geomZ + 2\rabiFrequency\,\cos(\radioFrequency \tim) \,\spin_\geomX + \gyro\,\groundTruthSignal(\tim) \,\spin_\geomZ,\label{eq:hamiltonian}
    \end{align}
    for duration $\sensingDuration$, before readout of the transverse spin projection (see Supplemental Material for details~\cite{Note0}). 
    The system approximately measures the Fourier sine coefficient of $\groundTruthSignal(\tim)$ at frequency $\frequency = \rabiFrequency/\mtau$, because the eigenvalues of its dressed frame 
    Hamiltonian are split by $\hhbar\,\rabiFrequency$:
    \begin{align}
      \hamiltonian^{\geomRot}(\tim) = \rabiFrequency \,\spin^{\geomRot}_\geomX + \gyro\,\groundTruthSignal(\tim) \,\spin_\geomZ \xrightarrow{\text{no signal}} \rabiFrequency \,\spin^{\geomRot}_\geomX.
    \end{align}
    The sine component is resonant with this splitting, driving transitions between dressed states at a rate proportional to $ \measurement(\frequency) $, \ie a shift of the dressed-frame plane of Rabi-flopping shown in \Figure~\ref{fig:overview}(b).
    Dressed state populations are mapped to lab frame $\spin_\geomZ$ populations $\atomsInMeasurement_{+,0,-}$ using a $\pi/2$ pulse [\acrshort*{rf} sequence in \Figure~\ref{fig:overview}(b)], which are then measured using Stern-Gerlach absorption imaging.
    As a result, we can measure different sine coefficients by repeating this, each time selecting a different \acrshort*{rf} amplitude $\rabiFrequency$.

  \begin{figure}[h!]
    \includegraphics[scale=0.65]{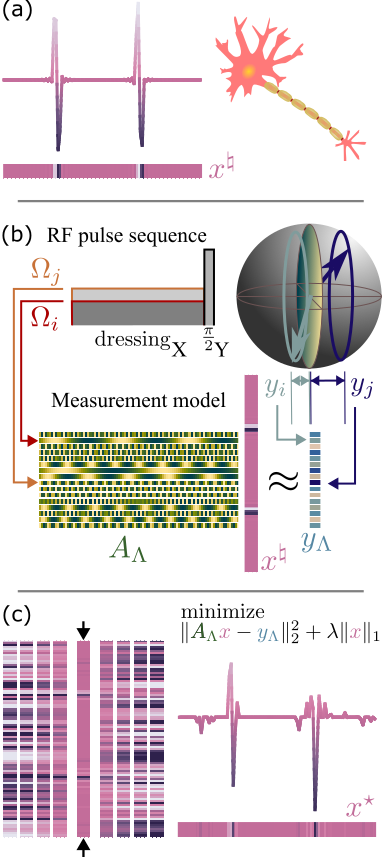}
    \caption[Compressive quantum waveform estimation]{
      Compressive quantum waveform estimation with \acrshort*{rf}-dressed atoms: 
      (a) Trapped atoms exposed to (synthes\is ed) neural (illustrated by \citet{phreed_single-neuron_2006} on right) magnetic signal $\groundTruthSignal$.
      Signal sampled in time-domain is illustrated as a col\ou r-coded vector underneath plot.
      (b) Two example pulse sequences (upper left) differing in their continuous dressing amplitude generate Rabi flopping at different Rabi frequencies $\rabiFrequency_{i, \frequencyIndex}$ depicted on Bloch sphere (upper right)\ox\ and as two rows of measurement matrix $\sampler_\subsampleIndexSet$ (lower left). Unitary evolution with weak signal $\groundTruthSignal$ and final measurement of transverse spin projection $\measurement_\frequencyIndex$ constitute sensing a \acrshort*{dst} coefficient for each $\rabiFrequency$, shown as linear system $\measurement_\subsampleIndexSet = \sampler_\subsampleIndexSet\signal$ (lower).
      Protocol mode\LL ed as an (exaggerated here) underdetermined matrix equation (lower).
      Equality is approximate due to measurement noise and simplification of model.
      Experiment is repeated for different \acrshort*{rf} amplitudes $\rabiFrequency_\frequencyIndex$ measuring different sine coefficients $\measurement_\frequencyIndex$.
      (c) \acrshort*{fista} searches the set of signals that fit the measurements (left) for the sparsest such signal, which it returns (right).
    }
    \label{fig:overview}
  \end{figure}
  \begin{figure*}[ht]
    \includegraphics[scale=0.55]{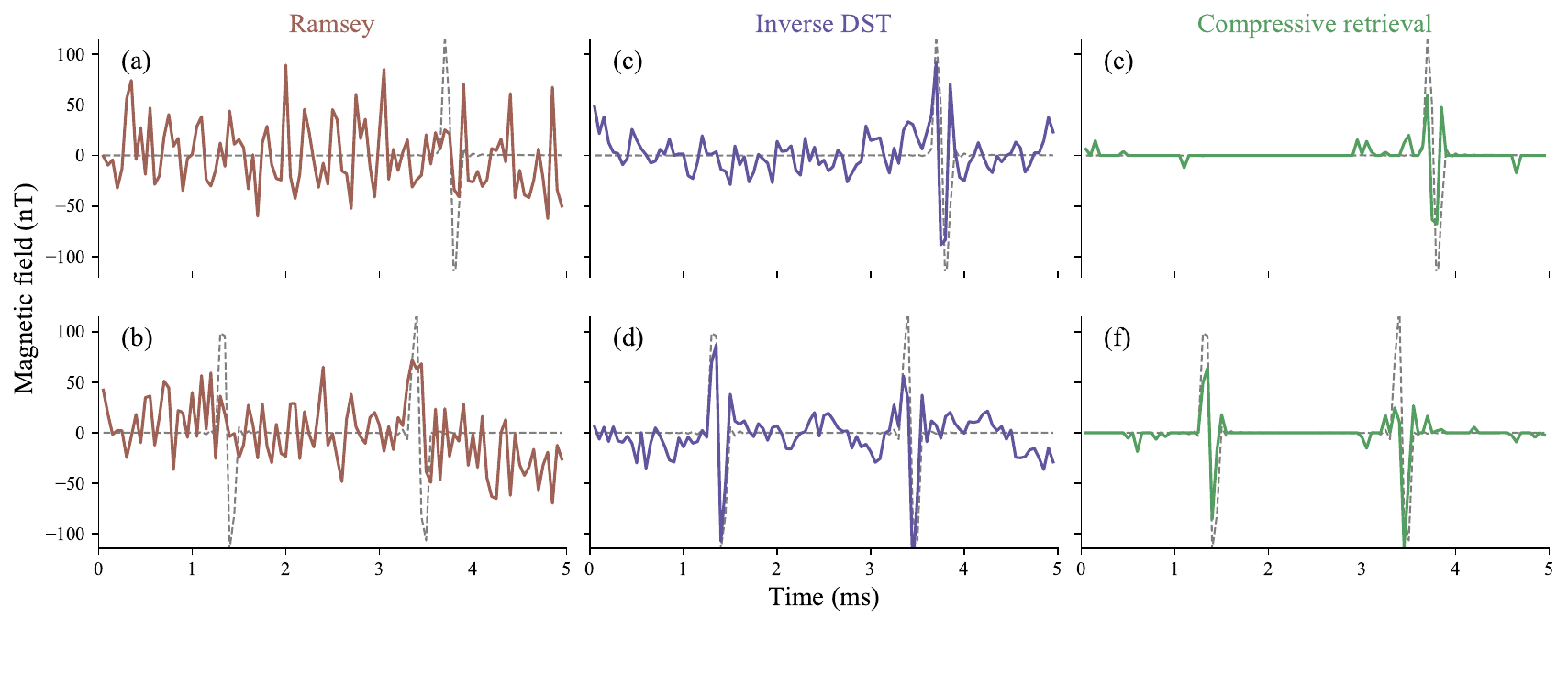}
    \caption[Neural-like magnetic pulses sensed in time domain, complete frequency domain and compressively.]{
      Neural-like magnetic pulses sensed in time domain, complete frequency domain and compressively.
      Solid line is recovered signal, dashed line is signal sent to coils.
      (a,b) Ramsey sampling of a (a) one- and (b) two-pulse signal.
      Signal is not able to be recovered due to bias drift.
      (c,d) Inverse \acrshort*{dst} using the complete set of sine coefficients recorded by the dressed atoms.
      (e,f) Compressive recovery using the \acrshort*{fista}. Only 60 of the 99 recorded sine coefficients are used.
    }
    \label{fig:comparison}
  \end{figure*}
  \prSection{Compressive recovery model}%
    We discret\is e our model for the measurement protocol in \Equation~\eqref{eq:dst} as
    \begin{align}
      \measurement_\frequencyIndex =& \frac{1}{\sensingDuration}\sum_{\timeIndex = 1}^{\sizeOfSignal - 1}\sin[\mtau (\frequencyIndex\,\diffLarge\frequency) (\timeIndex \,\diffLarge\tim)]\,\groundTruthSignal_\timeIndex\,\diffLarge\tim,\label{eq:descretised}
    \end{align}
    for $\frequencyIndex \in \{1, 2, \dots, \sizeOfSignal - 1\}$.
    Here the signal time series $\groundTruthSignal_\timeIndex$ starts at $\timeIndex = 1$ and has length $\sizeOfSignal - 1$.
    By choosing bandwidth as $\bandwidth \defi \sizeOfSignal\,\diffLarge\frequency = 1/(2\,\diffLarge\tim)$ and sensing duration $\sensingDuration \defi \sizeOfSignal \,\diffLarge\tim = 1/(2\,\diffLarge\frequency)$, \Equation~\eqref{eq:descretised} becomes the \acrshort*{dst}~\cite[\Equation~(37)]{jain_fast_1976};
    \ie the matrix equation $\measurement = \sampler\groundTruthSignal$, with $\sampler_{\frequencyIndex \timeIndex} = \sin(\pi \frequencyIndex\timeIndex/\sizeOfSignal)/\sizeOfSignal$.
    Compressive sensing theory roughly says that, since $\groundTruthSignal$ is sparse, and the \acrshort*{dct} is sensitive to all sparse signals, we can subsample our measurements and still be able to recover the signal~\cite[\Theorem~6]{koep_introduction_2019}: \ie measure only $\sizeOfMeasurement < \sizeOfSignal$ components $\measurement_\frequencyIndex$ of $\measurement$, so that now $\frequencyIndex \in \subsampleIndexSet \subset \{1, 2, \dots, \sizeOfSignal - 1\}$.
    \EQUATION~\eqref{eq:descretised} still holds, but only for $\frequencyIndex \in \subsampleIndexSet$; written $\measurement_\subsampleIndexSet = \sampler_\subsampleIndexSet \groundTruthSignal$, and illustrated as the matrix of sine waves of few randomly chosen frequencies in \Figure~\ref{fig:overview}(b).
    Therefore, we know that the signal must be one of the many solutions $\signal$ to the underdetermined system $\measurement_\subsampleIndexSet = \sampler_\subsampleIndexSet\signal$.
    
    This is not uniquely solvable, but because $\groundTruthSignal$ is sparse, the correct recovery, $ \reconstructedSignal $, will, with high probability, be the sparsest such $\signal$~\cite[\Theorem~6]{koep_introduction_2019}.
    In general this is an \acrshort*{np}-hard problem~\cite[\Theorem~2.17]{foucart_mathematical_2013}, but assuming that $\sampler_\subsampleIndexSet$ satisfies criteria involving the \acrfull*{rip}, the sparest solution of $\measurement_\subsampleIndexSet = \sampler_\subsampleIndexSet \groundTruthSignal$ will also be the solution which minim\is es the $\ell_1$-norm,
    $ \|\signal\|_1 \defi \sum_\timeIndex |\signal_\timeIndex| $~\cite[\Theorem~9]{koep_introduction_2019}.
    These criteria translate to requirements of measuring a system-dependent minimum number of sine coefficients~\cite[\Equation~(9.24)]{foucart_mathematical_2013} and for the subsampling of frequencies to have no structure (which is likely if they are chosen randomly~\cite[\Theorem~11.23]{foucart_mathematical_2013}).
    In practice, our linear model [\Equation~\eqref{eq:descretised}] will only be approximately satisfied due to measurement noise.
    Hence, we relax to a problem called the \emph{\acrfull*{lasso}}~\cite[\Equation~3.14]{santosa_linear_1986}, where we find the $\signal \in \realNumbers^{\sizeOfSignal - 1}$ that minim\is es $\|\sampler_\subsampleIndexSet \signal - \measurement_\subsampleIndexSet\|_2^2 + \regularisationParameter\|\signal\|_1$, illustrated in \Figure~\ref{fig:overview}(c).
    Here, the \emph{regular\is ation parameter} $ \regularisationParameter > 0 $~\cite[\Section~8.2.2]{koep_introduction_2019} is chosen~\cite{Note0} to produce the smallest error on simulated (using our open-source solver~\cite{tritt_spinsim_2023}) training data\ox\ and $\|\measurement_\subsampleIndexSet\|_2^2 \defi \sum_{\frequencyIndex\in\subsampleIndexSet} |\measurement_\frequencyIndex|^2$ is the $\ell_2$-norm.
    To solve the \acrshort*{lasso}, we implemented the \emph{\acrfull*{fista}}~\cite{beck_fast_2009}.
    We chose the step-size of \acrshort*{fista} using the singular values of $\sampler_\subsampleIndexSet$ and an an estimate of the amplitude of $\signal(\tim)$.

  \prSection{Experiment}\label{sec:experiment}%
    To experimentally verify this sensing method, we applied the Hamiltonian \Equation~\eqref{eq:hamiltonian} to a dipole-trapped cloud of approximately $6\mult10^5$ laser cooled atoms of $\rubidiumLxxxvii$ at a temperature of $1.0\unitSpace\mu\mathrm{K}$.
    A bias field of $86.1\unitSpace\mu\mathrm{T}$, gave a Larmor splitting of $\larmorFrequency = \mtau\mult603\unitSpace\mathrm{kHz}$, and we tuned our \acrshort*{rf} dressing $\radioFrequency$ to be resonant with this.
    The neural-like signal $\groundTruthSignal(\tim)$ comprised one or two pulses, each being a single cycle sine pulse of amplitude $143\unitSpace\mathrm{nT}$ and duration $200\unitSpace\mu\mathrm{s}$.
    The amplitude is much weaker than the $50\unitSpace\mu\mathrm{T}$ pulse used by \citet{webb_high-speed_2022}, though still stronger than sub-nanotesla fields expected in close proximity to individual neurons.
    The pulse duration is much longer than the $10\unitSpace\mu\mathrm{s}$ pulse simulated by \citet{magesan_compressing_2013}, though still shorter than the approximately $2\unitSpace\mathrm{ms}$ duration of a typical action potential.
    Starting in $\spinQuantumNumber = 1, \magneticQuantumNumber = -1$ (a superposition of dressed states), the sensor evolves under \Equation~\eqref{eq:hamiltonian}, before Zeeman populations $\atomsInMeasurement_{+,0,-}$ are recorded using Stern-Gerlach absorption imaging after a readout $\pi/2$ pulse.
    This is linearly sensitive to small $\measurement(\frequency)$ via $\measurement(\frequency) = \braket{\spin^{\geomRot}_{\geomX}}(\sensingDuration)/(\hhbar\,\gyro\,\sensingDuration) = \braket{\spin_\geomZ}_{\frac\pi2}/(\hhbar\,\gyro\,\sensingDuration)
    = (\atomsInMeasurement_- - \atomsInMeasurement_+)/[\gyro\,\sensingDuration\,(\atomsInMeasurement_- + \atomsInMeasurement_0 + \atomsInMeasurement_+)]$.
    
    The procedure was repeated, changing the value of $\rabiFrequency$ for each shot to record each frequency component from the incomplete set $\subsampleIndexSet$.
    Our mode\LL ed neural-like pulse had negligible sine components above $\bandwidth = 10\unitSpace\mathrm{kHz}$ (consistent with \citet{webb_detection_2021}) which fixed our highest meaningful time resolution to be $\diffLarge\tim = 50\unitSpace\mu\mathrm{s}$ (pulses 4 samples long).
    We chose an experiment duration of $\sensingDuration = 5\unitSpace\mathrm{ms}$ (99 samples long), yielding a frequency resolution of $\diffLarge\frequency = 100\unitSpace\mathrm{Hz}$.
    We do this to sense around the crest of a single cycle of the electrical line.

    We wished to compare measurements using our compressive protocol with complete measurements of both time and frequency samples.
    For the former, we used a separate Ramsey sequence to sample the amplitude of the signal at all of the 99 points on the time-grid.
    Here the second $\pi/2$ pulse was in quadrature to the first in order for the populations to be linearly sensitive to the signal (see the coherent signal case in \citet{mouradian_quantum_2021}).
    The Ramsey sequences approximate capturing the field at a fixed time by sensing over a rectangular window of length $60\unitSpace\mu\mathrm{s}$ cent\re d on the time-grid points.
    For the latter, we used the above protocol to measure a complete set of 99 sine coefficients and recovered using a complete inverse \acrshort*{dst}.
    We then obtained incomplete data $\measurement_\subsampleIndexSet$ for the compressive protocol by various subsamplings of these 99 sine coefficients $\measurement$.
    This enables us to directly compare complete inverse \acrshort*{dst}s with compressive retrievals.
    Of course, to benefit from compressive sensing, one would instead take fewer measurements in the first place.
    We present such experiments in Supplemental Material~\cite{Note0}.

    \begin{figure}[ht]
      \includegraphics[scale=0.55]{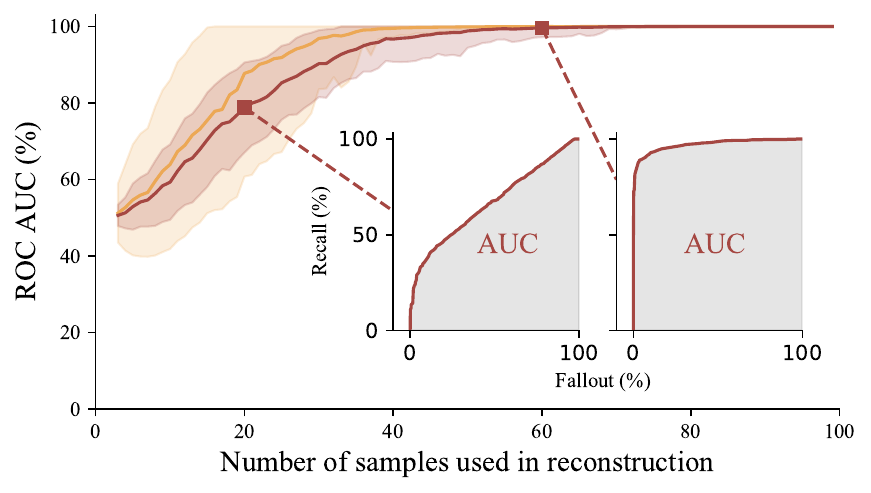}
      \caption[Finding the fewest number of samples that can be used by the compressive protocol before it fails in recovering the signal.]{
        Finding the fewest number of samples that can be used by the compressive protocol before it fails in recovering the signal.
        \acrshort*{auc} (see Supplemental Material~\cite{Note0}) of $100\unitSpaceSmall\%$ means perfect pulse detection, $50\unitSpaceSmall\%$ means random categor\is ation.
        \acrshort*{auc}s from 200 random subsets of the complete datasets were averaged to form each data point; shading shows one standard deviation either side.
        One-pulse traces (light orange) draw from sine coefficients used in \Figure~\ref{fig:comparison}(c), two-pulse traces (dark red) from \Figure~\ref{fig:comparison}(d).
        Left and right insets show (average over the 200 random subsets) \acrshort*{roc} curves for the two-pulse signal when 20 and 60 measurements are used in the recovery respectively.
        \acrshort*{roc}s are parameter\is ed by categor\is ation threshold.
        If a threshold perfectly distinguishes between pulses and noise, the \acrshort*{roc} reaches the top left and thus encloses an area (\acrshort*{auc} plotted in main figure) of $100\unitSpaceSmall\%$.
      }
      \label{fig:metrics}
    \end{figure}
    \prSection{Sensing results}%
    \FIGURE~\ref{fig:comparison} compares waveform measurements of repeated one- and two-pulse signals using the Ramsey protocol, an inverse \acrshort*{dst} of the complete set of sine coefficients, and the compressive protocol using a random set of 60 of the 99 sine coefficients.
    The Ramsey measurements in \Figures~\ref{fig:comparison}(a, b) are unable to detect the pulses.
    The recovery is overwhelmed by shot-to-shot drift of the bias magnetic field, which we later discovered was caused by suburban railway lines $2\unitSpace\mathrm{km}$ away.
    Our choice to shuffle the shot order made these slow drifts appear as white noise.
    Another source of deviation includes the interference from the \acrshort*{ac} electrical lines.
    Furthermore, the square sampling window of the Ramsey measurements makes the waveform estimation susceptible to high frequency noise from surrounding instrumentation.
    In contrast, the inverse \acrshort*{dst} retrieval in \Figures~\ref{fig:comparison}(c, d) were able to resolve the synthes\is ed pulses, as the \acrshort*{dst} successfully cuts out unwanted \acrshort*{dc} and high frequency coefficients.
    The compressive retrievals in \Figures~\ref{fig:comparison}(e, f), on the other hand, result in the cleanest signals of all the techniques, despite considering only 60 of the complete 99 sine coefficients.
    This is expected from the de-noising property of the \acrshort*{lasso}~\cite[\Figure~6(a)]{santosa_linear_1986}.
    
  \prSection{Compressive performance}\label{sec:compressive_performance}%
    Compressive sensing protocols have lower bounds as to how few samples can be used to recover a signal~\cite[\Equation~(9.24)]{foucart_mathematical_2013}.
    For our protocol, we wanted to experimentally determine this limit for our protocol and dataset.
    Here we compare our recovered signals to the waveform we commanded the magnetic coils with.
    While the \acrlong*{rmse} is a useful metric in other contexts, for our sparse signals we found that it had difficulty distinguishing noisy, but otherwise \emph{successful}, recoveries from completely failed recoveries.
    Canonical analysis of detecting signals in noise involves plotting a curve called the \acrfull*{roc}; the area under the curve (\acrshort*{auc})~\cite[\Section~7]{fawcett_introduction_2006} is an appropriate metric for determining signal recovery success~\cite{Note0}.
    A signal detector that never confuses signal and noise achieves an \acrshort*{auc} of $100\unitSpaceSmall\%$.
    In \Figure~\ref{fig:metrics}, the compressive protocol has \acrshort*{auc} over $ 99\unitSpaceSmall\% $ when at least $36$ samples are used to recover the signal containing one pulse, and at least $52$ samples for the signal containing two.
    For our parameters these echo the established theoretical bounds of $34$ and $57$ samples respectively~\cite[\Equation~(9.24)]{foucart_mathematical_2013}.
    We thus predicted that the protocol would work with $60$ coefficients measured, and it did, as shown in Supplemental Material~\cite{Note0}.
    While a reduction in quantum resources by almost half is already useful, the theoretical model predicts even larger relative gains when measuring longer, sparser signals.
  
  \prSection{Conclusion}%
    The compressive quantum sensor was able to recover neural-like signals in a situation where Ramsey measurements could not, while using an incomplete set of measurements.
    Choice of signal basis can drastically change both the fidelity, and, when undersampling, efficiency of quantum waveform estimation.
    While the implementation presented in this Letter arrays measurement in time to estimate a repeated waveform, we envision an implementation on a quantum sensing array of independently-controlled sensors arrayed in space, where resource parsimony is of the essence. 
    Such an array would be capable of compressive waveform estimation of a never-to-be-repeated signal, in a single shot. 

    \prSection{Acknowledgements}%
    We thank Anton van den Hengel, Kristian Helmerson and David Spanswick for helpful discussions.
    AT, CCB\ox\ and HAMT acknowledge support through \fundRtp s.
    JS is the recipient of an \fundJamesZero\ funded by the Australian Government.
    LDT acknowledges funding from the \fundLincolnZero. 
  \bibliography{paper}
  
  \fi
  \if\makeSupMat1
  \widetext
  \clearpage
  
  \setcounter{page}{1}
  \begin{center}
    \textbf{\large Supplemental Material}
  \end{center}

  \section{Protocol for sensing Fourier sine coefficients}\label{sec:derivation}
    The lab frame Hamiltonian of our atoms is
    \begin{align}
      \hamiltonian(\tim) = \larmorFrequency \,\spin_\geomZ + 2\rabiFrequency\,\cos(\radioFrequency\tim) \,\spin_\geomX + \gyro\,\groundTruthSignal(\tim) \,\spin_\geomZ.
    \end{align}
    Moving into a frame $\geomRot$ rotating at $\radioFrequency = \larmorFrequency$ (on resonance) around $\spin_\geomZ$, and taking the \acrfull*{rwa}, the system has a Hamiltonian of
    \begin{align}
      \hamiltonian^{\geomRot}(\tim) = \rabiFrequency \,\spin^{\geomRot}_\geomX + \gyro\,\groundTruthSignal(\tim) \,\spin_\geomZ.
    \end{align}
    We can now move into a second rotating frame $\geomRot\geomRot$ at $\rabiFrequency$ around $\spin^{\geomRot}_\geomX$, but without taking a second \acrshort*{rwa}:
    \begin{align}
      \hamiltonian^{\geomRot\geomRot}(\tim) = \gyro\,\groundTruthSignal(\tim)\left(\sin(\rabiFrequency\,\tim)\,\spin^{\geomRot\geomRot}_\geomY - \cos(\rabiFrequency\,\tim)\,\spin^{\geomRot\geomRot}_\geomZ\right).
    \end{align}

    We assume that the signal $\groundTruthSignal$ is always small enough such that time evolution can be well approximated by a truncated Magnus series $\sum_\magnusIndex\magnusOperator_\magnusIndex$~\cite{blanes_magnus_2009}.
    Then the time evolution operator of the spin in the second rotating frame over the sensing duration $\sensingDuration$ is
    \begin{align}
      \unit =& \exp\left(\magnusOperator_1 + \magnusOperator_2+ \magnusOperator_3 + \cdots\right),\\
      \unit_1 =& \exp\left(\magnusOperator_1\right)\quad\text{to first order},\\
      =& \exp\left(\frac{i\,\gyro}{\hhbar}\int_{0}^{\sensingDuration} \groundTruthSignal(\tim)\left(\sin(\rabiFrequency\,\tim)\,\spin^{\geomRot\geomRot}_\geomY - \cos(\rabiFrequency\,\tim)\,\spin^{\geomRot\geomRot}_\geomZ\right)\diff\tim\right)\quad\text{from \cite[\Equation~(42)]{blanes_magnus_2009}},\\
      =& \exp\left(\frac{i\,\gyro}{\hhbar}\left(\int_{0}^{\sensingDuration} \sin(\rabiFrequency\,\tim)\,\groundTruthSignal(\tim)\,\diff\tim\,\spin^{\geomRot\geomRot}_\geomY - \int_{0}^{\sensingDuration}\cos(\rabiFrequency\,\tim)\,\groundTruthSignal(\tim)\,\diff\tim\,\spin^{\geomRot\geomRot}_\geomZ\right)\right),\\
      =& \exp\left(\frac{i}{\hhbar}\left(\sinCoef\,\spin^{\geomRot\geomRot}_\geomY - \cosCoef\,\spin^{\geomRot\geomRot}_\geomZ\right)\right).
    \end{align}
    Here $\sinCoef = \gyro\int_{0}^{\sensingDuration} \sin(\rabiFrequency\,\tim)\,\groundTruthSignal(\tim)\,\diff\tim = \gyro\,\sensingDuration\,\measurement(\rabiFrequency/\mtau)$ is proportional to the Fourier sine coefficient $\measurement(\rabiFrequency/\mtau)$ of the signal $\groundTruthSignal$ at frequency $\rabiFrequency/\mtau$, which is the quantity we want to measure.
    Similarly $\cosCoef = \gyro\int_{0}^{\sensingDuration}\cos(\rabiFrequency\,\tim)\,\groundTruthSignal(\tim)\,\diff\tim$ is proportional to the Fourier  cosine coefficient of $\groundTruthSignal$ at frequency $\frequency = \rabiFrequency/\mtau$.
    Note that the second term in the Magnus expansion is~\cite[\Equation~(43)]{blanes_magnus_2009}
    \begin{align}
      \magnusOperator_2 &= \frac{i\,\gyro^2}{2\,\hhbar}\int_{0}^{\sensingDuration}\int_{0}^{\tim_1}\groundTruthSignal(\tim_1)\,\groundTruthSignal(\tim_2)\,\sin\left(\rabiFrequency\,(\tim_1 - \tim_2)\right)\,\diff\tim_2\,\diff\tim_1\,\spin^{\geomRot\geomRot}_\geomX,
    \end{align}
    which is suppressed by being of order $(\groundTruthSignal)^2$\ox\ and because it drives rotations in the same axis as the dressing in the first rotating frame.
    Hence we only consider the dynamics due to $\unit_1$.

    We show later [in \Equation~\eqref{eq:population_count}] that we can measure $\braket{\spin^{\geomRot\geomRot}_{\geomX}(\sensingDuration)} = \bra{\wavefunction(\sensingDuration)}\spin^{\geomRot\geomRot}_{\geomX}\ket{\wavefunction(\sensingDuration)} = \bra{\wavefunction(0)}\,\unit_1^\dagger\,\spin^{\geomRot\geomRot}_{\geomX}\,\unit_1\ket{\wavefunction(0)}$ for an experiment starting in some initial state $\ket{\wavefunction(0)}$.
    First, note that the unitary $\unit_1$ is a rotation on the Bloch sphere of angle $\sqrt{\sinCoef^2 + \cosCoef^2}$ around an axis lying on the $\geomY\geomZ$ plane at an angle $\mathrm{arctan2}\left(\sinCoef, \cosCoef\right)$ to the $\geomZ$ axis.
    Second, note that the operator $\spin^{\geomRot\geomRot}_{\geomX}$ we are taking an expectation value of is a component of a vector operator $\mathbf{\spin}^{\geomRot\geomRot}$.
    Therefore, we can simply evaluate the similarity transform $\unit_1^\dagger\,\spin^{\geomRot\geomRot}_{\geomX}\,\unit_1$ in this expression using standard 3-vector rotation matrices~\cite[\Equation~(3.10.3)]{j_j_sakurai_jun_john_modern_1994}.
    In fact, the general result of \Equation~(121) from \citet{siminovitch_exact_2003} allows us to readily evaluate the similarity transform as
    \begin{align}
      \unit_1^\dagger\,\spin^{\geomRot\geomRot}_{\geomX}\,\unit_1 &= - \frac{\sin\left(\sqrt{\sinCoef^2 + \cosCoef^2}\right)}{\sqrt{\sinCoef^2 + \cosCoef^2}}\,\sinCoef\,\spin^{\geomRot\geomRot}_{\geomX} - \frac{\sin\left(\sqrt{\sinCoef^2 + \cosCoef^2}\right)}{\sqrt{\sinCoef^2 + \cosCoef^2}}\,\cosCoef\,\spin^{\geomRot\geomRot}_{\geomY} + \cos\left(\sqrt{\sinCoef^2 + \cosCoef^2}\right)\,\spin^{\geomRot\geomRot}_{\geomZ}
    \end{align}

    We can then trivially read-off the expectation value $\braket{\spin^{\geomRot\geomRot}_{\geomX}}(\sensingDuration)$ when our initial state is an eigenstate of each of the spin operators.
    In particular, when starting in eigenstate $\ket{-\geomZ}$ of $\spin^{\geomRot\geomRot}_{\geomZ}$, the expectation value is 
    \begin{align}
      \braket{\spin^{\geomRot\geomRot}_{\geomX}}(\sensingDuration)/\hhbar = \frac{\sin\left(\sqrt{\sinCoef^2 + \cosCoef^2}\right)}{\sqrt{\sinCoef^2 + \cosCoef^2}}\,\sinCoef.
    \end{align}
    That is (continuing with the assumption that $\groundTruthSignal$ and hence $\sqrt{\sinCoef^2 + \cosCoef^2}$ are small), when starting in $\ket{-\geomZ}$, the expectation value is linearly sensitive to the Fourier sine coefficient of $\groundTruthSignal$ that we wish to measure.
    Therefore we choose to start in $\ket{-\geomZ}$.
    Note that, although not used in our protocol, when starting in eigenstate $\ket{+\geomY}$ of $\spin^{\geomRot\geomRot}_{\geomY}$ we are similarly linearly sensitive to the cosine coefficient of $\groundTruthSignal$ at frequency $\frequency = \rabiFrequency/\mtau$\ox\ and when starting in $\ket{+\geomX}$ of $\spin^{\geomRot\geomRot}_{\geomX}$ we are similarly linearly sensitive to the spectral power (and quadratically sensitive to the magnitude of the complex Fourier component) of $\groundTruthSignal$ at frequency $\frequency = \rabiFrequency/\mtau$.

    At this point the analysis of the measurement protocol is entirely in the second rotating frame.
    First note that a state that starts in an eigenstate of $\spin_{\geomZ}$ in the lab frame will also start in an eigenstate of $\spin^{\geomRot\geomRot}_{\geomZ}$ in the second rotating frame.
    Thus we prepare our atoms in the lab frame $\ket{-\geomZ}$ state, which is conveniently the state that the magnetic catch cooling stage polar\is es our atoms into.
    We then let these atoms evolve such that they have an expected $\braket{\spin^{\geomRot\geomRot}_{\geomX}}(\sensingDuration)$ linearly sensitive to the Fourier sine coefficient.
    Note that $\braket{\spin^{\geomRot\geomRot}_{\geomX}}(\sensingDuration)$ in the second rotating frame is the same as $\braket{\spin^{\geomRot}_{\geomX}}(\sensingDuration)$ in the first rotating frame.
    We finish the pulse sequence by applying a $20\unitSpace\text{kHz}$ $\pi/2$ read-out pulse around $\spin^{\geomRot}_\geomY$ to rotate this expected value from $\spin^{\geomRot}_\geomX$ to $\spin_\geomZ$.
    The populations of the eigenstates of $\spin_\geomZ$ are then imaged using a Stern-Gerlach projective measurement.
    Last, we record our sine coefficients as 
    \begin{align}
      \measurement(\frequency) &= \frac{\braket{\spin^{\geomRot}_{\geomX}}(\sensingDuration)}{\hhbar\,\gyro\,\sensingDuration}\\
      &= \frac{\braket{\spin_\geomZ}_{\frac\pi2}}{\hhbar\,\gyro\,\sensingDuration},\\
      &= \frac{1}{\gyro\,\sensingDuration}\frac{\atomsInMeasurement_+ - \atomsInMeasurement_-}{\atomsInMeasurement_+ + \atomsInMeasurement_0 + \atomsInMeasurement_-}.\label{eq:population_count}
    \end{align}

  \section{Tuning of regular\is ation parameter}\label{sec:regularisation}
    The \acrshort*{lasso} problem needs to be tuned by the regular\is ation parameter $\regularisationParameter$~\cite[\Section~8.2.2]{koep_introduction_2019}, based on the typical noise and sparsity of measurements.
    To do this, we used mathematical optim\is ation to find a $\regularisationParameter$ that produced the smallest average error over a set of one-thousand measurements of signals.  
    Since collecting such data manually would require on the order of a month of time, we instead used our open-source \acrshort*{gpu} accelerated simulation python package \emph{spinsim}~\cite{tritt_spinsim_2023} to generate a set of simulated experiment results.
    Each simulated sequence was randomly assigned 0, 1 or 2 neural-like pulses of amplitude $1\unitSpace\mathrm{kHz}/\gyro$ (\ie $143\unitSpace\mathrm{nT}$), each randomly placed in time.
    Imperfections were mode\LL ed in these experiments in the form of shot-to-shot bias drift (standard deviation of $200\unitSpace\mathrm{Hz}/\gyro$ consistent with lab measurements) and Poisson-distributed atom shot noise (expected number of atoms depreciated from order million to one thousand to also take into account photon shot noise).
    The error metric we chose was the average $\ell_1$-norm of the difference between true and measured signals.
    This is the middle-ground between the $\ell_0$-norm, $\|\signal\|_0\defi \#\{\timeIndex \text{ such that } \signal_\timeIndex = 0\}$~\cite[\Section~2.1]{koep_introduction_2019} which fav\ou rs correct supports, and the $\ell_2$-norm which fav\ou rs correct amplitudes.
    Note that we decided \emph{against} using \acrfull*{auc} (see below) as a metric for training since its landscape is very flat and therefore not informative for training.
    We scanned $\regularisationParameter$ over 200 different values with log spacing from $100\unitSpace\mathrm{mHz}/\gyro$ to $10\unitSpace\mathrm{Hz}/\gyro$, and found that value that returned the smallest error was $\regularisationParameter = 1.04\unitSpace\mathrm{Hz}/\gyro$.
    We emphas\is e that tuning $\regularisationParameter$ this precisely is not necessary\ox\ and using $\regularisationParameter$ in the range increased the $\ell_1$ norm by less than $10\unitSpaceSmall\%$ and yielded quality recoveries.

  \section{A signal detection statistic: the area under the receiver operating characteristic (ROC) curve (AUC)}\label{sec:metrics}
    There are three kinds of signals of interest when anal\ys ing acquired synthes\is ed neural-like pulses:
    We have the ground-truth signal $\groundTruthSignal$, which is a recording of the electrical signal commanded to the magnetic coils.
    We have the recovered signal $\reconstructedSignal$, which is the result of the compressive reconstructions or the Ramsey measurements.
    Finally, we have the pulse template $\pulseTemplate$, which is the expected shape of each individual pulse, but it starts at $\tim = 0$.

    The highest fidelity linear method of detecting pulse locations within a signal is by using a matched filter~\cite[\Section~4.3.1]{kay_fundamentals_1998}.
    In a matched filter, the template $\pulseTemplate$ is translated to all the possible starting locations in the signal, and an inner-product is taken between the two via $\groundTruthMatchedOutput_\timeIndex = \sum_\frequencyIndex \groundTruthSignal_\frequencyIndex\,\pulseTemplate_{\frequencyIndex + \timeIndex}$.
    This can be written succinctly using the cross-correlation operator via $\groundTruthMatchedOutput = \groundTruthSignal\crossCorrelation \pulseTemplate$, $\reconstructedMatchedOutput = \reconstructedSignal\crossCorrelation \pulseTemplate$.
    A high value of $\groundTruthMatchedOutput_\timeIndex$ or $\reconstructedMatchedOutput_\timeIndex$ at any index corresponds to a high likelihood that a pulse is located in $\groundTruthSignal$ or $\reconstructedSignal$ respectively at time index $\timeIndex$.
    We can classify locations as having a pulse or not by thresholding $\groundTruthMatchedOutput$ and $\reconstructedMatchedOutput$.
    We obtain a ground truth classification $\groundTruthClassification = \decisionThresholder_{\|\pulseTemplate\|_2^2/2}(\groundTruthMatchedOutput)$ by taking a threshold of the ground truth signal at a level of $\|\pulseTemplate\|_2^2/2$.
    Here we define $\decisionThresholder_{\threshold}(\groundTruthMatchedOutput)_\timeIndex = 0$ for $\groundTruthMatchedOutput_\timeIndex < \threshold$, and $\decisionThresholder_{\threshold}(\groundTruthMatchedOutput)_\timeIndex = 1$ otherwise.
    
    Classifications can be made of the recovered signal $\reconstructedClassification = \decisionThresholder_{\threshold}(\reconstructedMatchedOutput)$ at various thresholds $\threshold$.
    For each $\threshold$, we can obtain the statistics of a \emph{confusion matrix}~\citet[\Figure~1]{fawcett_introduction_2006} by comparing the classifications of individual locations in $\reconstructedClassification$ to those in $\groundTruthClassification$,
    \begin{align}
      \text{\# True Positives} = \truePositives =& \sum_\timeIndex \reconstructedClassification_\timeIndex \groundTruthClassification_\timeIndex, \\
      \text{\# False Positives} = \falsePositives =& \sum_\timeIndex \reconstructedClassification_\timeIndex (1 - \groundTruthClassification_\timeIndex), \\
      \text{\# False Negatives} = \falseNegatives =& \sum_\timeIndex (1 - \reconstructedClassification_\timeIndex) \groundTruthClassification_\timeIndex,\quad\text{and} \\
      \text{\# True Negatives} = \trueNegatives =& \sum_\timeIndex (1 - \reconstructedClassification_\timeIndex) (1 - \groundTruthClassification_\timeIndex).
    \end{align}
    
    Two important statistics~\cite[\Section~2]{fawcett_introduction_2006} can be derived from these values.
    The recall (which we write $\recall$, also sensitivity, true-positive rate or hit rate in the literature) is defined as
    \begin{align}
      \recall \defi& \frac{\truePositives}{\truePositives + \falseNegatives},
    \end{align}
    and is the fraction of ground-truth positives correctly detected in the recovery.

    Similarly, fallout (which we write $\fallout$, also false-positive rate, false-alarm rate or $1 - \text{specificity}$ in the literature) is defined as
    \begin{align}
      \fallout \defi& \frac{\falsePositives}{\falsePositives + \trueNegatives},
    \end{align}
    and is the fraction of ground-truth negatives incorrectly detected in the recovery.
    Note that these are both functions of the recovery threshold $\threshold$.
    The parametric curve of points $\left(\fallout(\threshold), \recall(\threshold)\right)$ that $\threshold$ traces out is called the \acrfull*{roc} of the sensor~\cite[\Section~4]{fawcett_introduction_2006}.
    When $\threshold\to-\infty$, any noise will be classified as a signal, meaning $(\fallout(\threshold), \recall(\threshold))\to(1, 1)$.
    When $\threshold\to\infty$, none of the true signal responses will be large enough to be classified as a signal, so $(\fallout(\threshold),\recall(\threshold))\to(0, 0)$.
    If there is some threshold $\thresholdPerfect$ that can completely accurately distinguish between pulses and noise, then $(\fallout(\threshold), \recall(\threshold)) = (0, 1)$.
    
    The closer the \acrshort*{roc} curve gets to this value of $(0, 1)$, the better the classifier.
    This can be quantified by integrating the \acrfull*{auc}~\cite[\Section~7]{fawcett_introduction_2006}, $\auc$.
    If the \acrshort*{roc} curve crosses $(0, 1)$, then the area is a square and $\auc = 1$. %
    If the classifier does no better than randomly guessing, then the area is a triangle under $\fallout = \recall$ and $\auc = 1/2$.
    
    \FIGURE~\ref{fig:metrics} in the \Letter\ uses \acrshort*{auc} as a quality metric of compressive recoveries.
    The insets of the figure show examples of \acrshort*{roc}s for recoveries made using different compression ratios, averaged for 200 different subsamples.
    We found that, for our case of using data from only one signal, the number of subsamples where the average \acrshort*{auc} drops below $100\unitSpaceSmall\%$ matches that where we would expect failed recoveries in compressive sensing theoretical guarantees~\cite[\Equation~(9.24)]{foucart_mathematical_2013}.

  \section{Genuinely undersampled compressive measurements}\label{sec:unknown}
    \begin{figure}[ht]
      \includegraphics[scale=0.55]{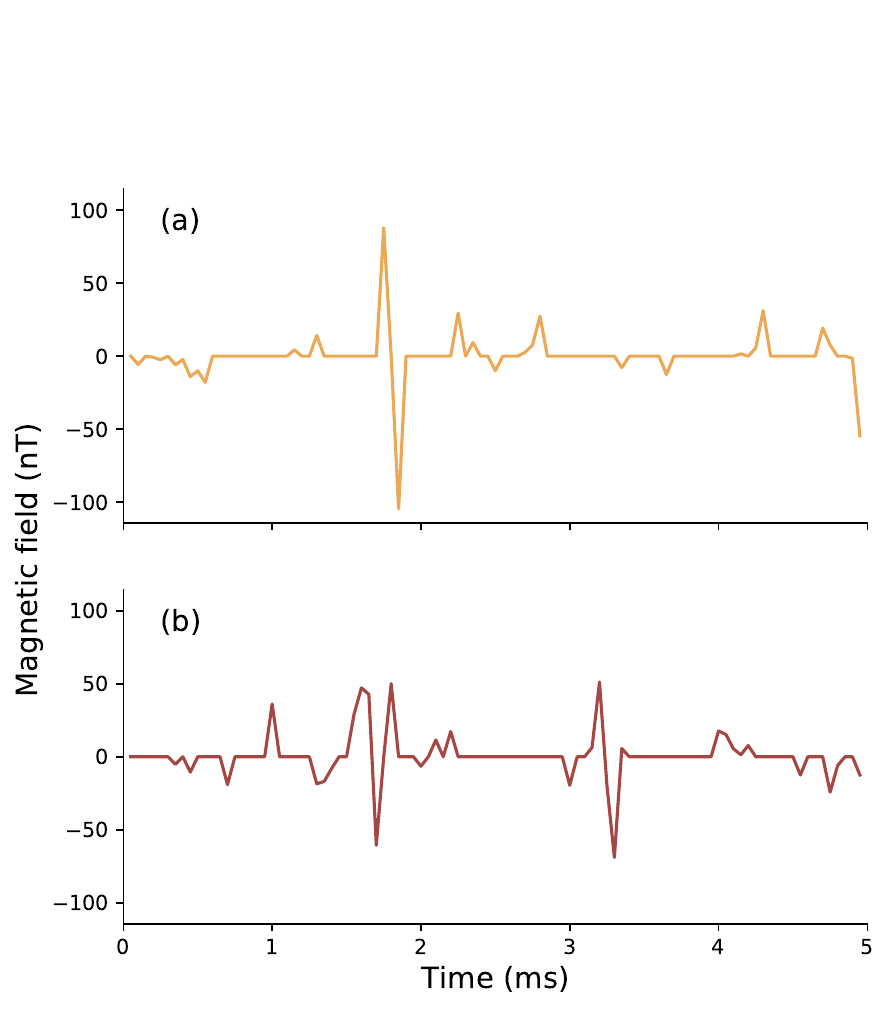}
      \caption[Recoveries of two genuinely undersampled signals.]{
        Recoveries of two genuinely undersampled signals, each with only of the 60 of the complete set of 99 sine coefficients measured by the cold atoms.
        (a) a single pulse signal recovery.
        (b) a two-pulse signal recovery
      }
      \label{fig:unknown}
    \end{figure}
    While using the atoms to sample all sine coefficients is necessary to analyse the effectiveness of quantum compressive sensing, recording a complete dataset misses the spirit of the kind of sensor we demonstrate.
    In this spirit, we show compressive recovery of signals from incomplete measurements of sine coefficients.

    We chose 60 random frequencies out of the complete \acrshort*{dst} (\ie multiples of $100\unitSpace\mathrm{Hz}$).
    We used the atoms to measure sine coefficients of a synthes\is ed waveform at these 60 frequencies, and only these frequencies.
    \acrshort*{fista} then recovered the waveform based on this incomplete information.

    \FIGURE~\ref{fig:unknown} shows compressive waveform estimation from genuinely incomplete data sets.
    It is evident on inspection that, as per the signal detection analysis in the previous \Section, there is a range of thresholds that will achieve correct detection of pulses.
  \if\makeLetter0
  \bibliography{paper}
  \fi
  \fi
\end{document}